\documentclass[%
 reprint,
 superscriptaddress,
 amsmath,amssymb,
 aps,
]{revtex4-2}

\usepackage{graphicx}% Include figure files
\usepackage{dcolumn}% Align table columns on decimal point
\usepackage{bm}% bold math
\usepackage{hyperref}% add hypertext capabilities
\usepackage{color}%追加
\usepackage{comment}%追加

\DeclareUnicodeCharacter{2061}{}
\usepackage{textcomp}
\usepackage{mathcomp}

\begin{document}

%\preprint{APS/123-QED}

\title{Frequency-multiplexed Hong-Ou-Mandel interference}% Force line breaks with \\

\author{Mayuka Ichihara}
\email{ichihara-mayuka-wm@ynu.jp}
\affiliation{
Department of physics, Yokohama National University, 79-5 Tokiwadai, Hodogaya, Yokohama 240-8501, Japan
}

\author{Daisuke Yoshida}
\affiliation{
Department of physics, Yokohama National University, 79-5 Tokiwadai, Hodogaya, Yokohama 240-8501, Japan
}
\affiliation{
LQUOM Inc., 79-5 Tokiwadai, Hodogaya, Yokohama 240-8501, Japan
}

\author{Feng-Lei Hong}
\affiliation{
Department of physics, Yokohama National University, 79-5 Tokiwadai, Hodogaya, Yokohama 240-8501, Japan
}
\author{Tomoyuki Horikiri}
\affiliation{
Department of physics, Yokohama National University, 79-5 Tokiwadai, Hodogaya, Yokohama 240-8501, Japan
}
\affiliation{
LQUOM Inc., 79-5 Tokiwadai, Hodogaya, Yokohama 240-8501, Japan
}

\date{\today}

\begin{abstract}
The implementation of quantum repeaters needed for long-distance quantum communication requires the generation of quantum entanglement distributed among the elementary links. These entanglements must be swapped among the quantum repeaters through Bell-state measurements. This study aims to improve the entanglement generation rate by frequency multiplexing the Bell-state measurements. As a preliminary step of the frequency-multiplexed Bell-state measurements, three frequency modes are mapped to a temporal mode by an atomic frequency comb prepared in $\mathrm{Pr^{3+}}$ ion-doped $\mathrm{Y_2SiO_5}$ crystals using a weak coherent state, and Hong-Ou-Mandel interference, which is a measure of the indistinguishability of two inputs, is observed in each frequency mode by coincidence detection. The visibility for all the modes was 40\%--42\% (theoretically up to 50\%). Furthermore, we show that a mixture of different modes is avoided. The present results are connected to frequency-selective Bell-state measurements and therefore frequency-multiplexed quantum repeaters.
\end{abstract}

\maketitle

%\tableofcontents
%%%%%%%イントロ！！！！%%%%%%%
\section{\label{level1-1}INTRODUCTION}

Quantum communication is a method that has gained significant attention owing to its potential applications including quantum key distribution (QKD) \cite{Nicolas}, distributed quantum computation \cite{Cirac}, and blind quantum computation \cite{Anne,Barz,Huang}. The ability to share entanglement and sending quantum states among remote users are essential parts of implementing distributed quantum computing and blind quantum computing. 

\begin{figure}
    \centering
    \includegraphics[width = 8.6cm]{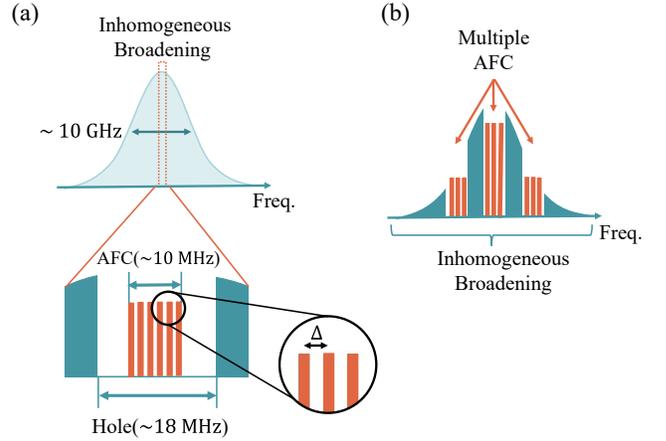}
    \caption{
    AFC prepared in Pr:YSO. (a) An 18-MHz hole is generated in the inhomogeneous broadening (approximately 10 GHz), and comb like absorption peaks, AFCs (approximately 10 MHz), are created in it. (b) Multiple AFCs in inhomogeneous broadening.}
    \label{fig:Pr:YSO}
\end{figure}

In general, it is difficult to transmit quantum states over long distances because of the loss and/or decoherence of the transmitted quantum states. The quantum repeater, proposed by Ref. \cite{Briegel}, is a solution to this problem, which enables entangled photons to be shared over long distances; it is an essential mechanism in the realization of the quantum internet \cite{Kimble,Wehner}. The components necessary for entanglement swapping, which is the function of a quantum repeater, include quantum memories (QMs) and Bell-state measurement (BSM). BSM is a measurement that projects to the Bell state, which allows entanglement swapping. The QM plays the role of synchronizing entanglement swapping for the generation of entanglement between non-neighboring nodes. Examples of QMs include rare-earth-doped materials like $\mathrm{Pr^{3+}}$ ion-doped $\mathrm{Y_2SiO_5}$ (Pr:YSO) \cite{Riedmatten,Afzelius}, Eu:YSO \cite{Ortu20ms}, Er-doped fibers \cite{Erhan}, Nd:$\mathrm{YVO_4}$ \cite{Xiao}, atomic gases \cite{Julsgaard}, and diamond nitrogen-vacancy centers \cite{Doherty,Pompili}. The atomic frequency comb (AFC) \cite{Afzelius}, in which an inhomogeneous broadening of rare-earth-doped material is controlled to provide a memory function, has the advantage of being multimode storage. Pr:YSO has relatively long coherence times, high multiplicities, and high efficiencies \cite{Dario}. Therefore, in the present work, we focus on Pr:YSO. AFCs are generated by hole burning in an inhomogeneous broadening of Pr:YSO (Fig. \ref{fig:Pr:YSO}).

AFCs can store photons for a duration equal to the reciprocal of the comb interval $\Delta$. We use hyperfine levels within the crystal field splitting between $\mathrm{^{3}H_{4}}$ and $\mathrm{^{1}D_{2}}$. Photons stored at $t = 0$ are retrieved as photon echoes at $t = 1/\Delta$. Because the inhomogeneous broadening is approximately 10 GHz \cite{Equall} wide and the bandwidth of the AFC is less than 10 MHz, it is possible to create multiple AFCs in the inhomogeneous broadening. Therefore, researchers have already carried out studies using multiplexing with AFCs \cite{Seri,Sinclair}. Frequency multiplexing is expected to improve the entanglement generation rate by adopting frequency-multiplexed QMs and frequency-selective optical BSMs \cite{Sinclair}. This is discussed in Sec. II. 

BSM requires indistinguishable photon pairs, and especially in the case of linear optical BSM, Hong-Ou-Mandel (HOM) interference \cite{HOM} is required at the beam splitter (BS). Therefore, in this study, we observe HOM interference. HOM interference is a phenomenon in which when an indistinguishable photon pair enters the 50:50 BS, only one-sided detection occurs. For the dip obtained through this HOM interference, the visibility $V$ is defined as follows:
\begin{equation}
V=\frac{P_{max}-P_{min}}{P_{max}}
\end{equation}
where $P_{max}$ and $P_{min}$ are the maximum and minimum values of the probability of coincidence detection in the two detectors. In HOM interference using a single photon, $V\sim1$ \cite{HOM}, but with a weak coherent state, it is known that $V\leqq0.5$ due to the Poisson distribution of the laser \cite{Ghosh}.

To implement frequency-multiplexed BSM for entanglement generation between elementary links, although earlier studies achieved HOM interference with one-mode photon echoes \cite{Jeongwan} and frequency-multiplexed HOM interference with frequency-space mapping \cite{Oriol}, there are no reports of frequency-multiplexed HOM experiments using multiple AFCs. In this study, we report on the preparation of AFCs with three-mode frequency multiplexing in each of two Pr:YSOs and the observation of HOM interference in each of the three modes. It is expected that the combination of frequency-time mapping and frequency-space mapping will increase the extent of multiplexing and lead to higher entanglement generation rates. 

The rest of this paper is organized as follows. The scheme of the frequency-multiplexed BSM is presented in Sec. II. The experimental setup is described in Sec. III. The results and discussion are presented in Sec. IV, and finally, the conclusion is presented in Sec. V.

%%%%%%%スキーム！！！！%%%%%%%
\section{\label{level1-2}SCHEME OF FREQUENCY MULTIPLEXED BSM AND INCREASE OF THE ENTANGLEMENT GENERATION RATE}
\begin{figure}
    \centering
    \includegraphics[width = 8.6cm]{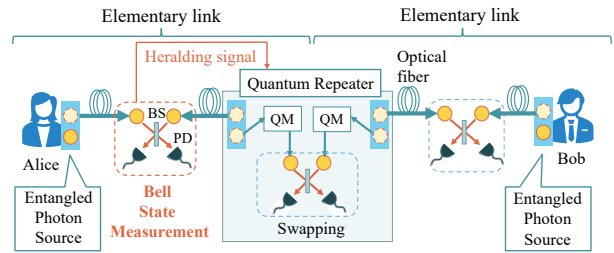}
    \caption{
    Schematic diagram of the quantum repeater, where the successful signal of the Bell-state measurement is a heralding signal, allowing the repeater to perform entanglement swapping using the quantum memories (QMs). In this study, we assume that the BSMs outside the repeater are multiplexed.}
    \label{fig:scheme}
\end{figure}

In this section, we will describe the adopted schemes and the estimation of how much the entanglement generation rate can be improved by frequency multiplexing. The scheme is illustrated in Fig. \ref{fig:scheme} \cite{Sinclair,Cody}, where there is one quantum repeater and we assume that the QMs can store for only a fixed amount of time, following Ref. \cite{Sinclair}. 

There are various schemes for quantum repeaters. Herein, what we assume in this study is the repeater node. There are two typical entanglement distribution schemes between the end points of each elementary link of the entire quantum repeater architecture, as described in Ref. \cite{Cody}: one is called ``meet in the middle,'' in which a quantum memory and an entangled photon source are installed in the repeater node. Photons emitted from the left and right repeater nodes enter an optical BSM at the midpoint, which becomes a heralding signal that generates entanglement between end points of each elementary link. The other scheme is called ``midpoint source,'' in which a quantum memory is installed in the repeater node and an entangled photon source is installed at the midpoint of an elementary link. In this method, it is difficult to use an absorptive quantum memory. The reason is because it is not possible for an absorptive quantum memory to generate a heralding signal that indicates that the entangled photons coming from the midpoint have been absorbed and stored in the quantum memory. Thus, it is necessary to implement a new nondestructive detection function for photon arrival just before the quantum memory, but this technology is not mature at present. In contrast, for the meet-in-the-middle case, the signal from the optical BSM device at the midpoint can be used as the heralding signal. This study will lead to frequency-channel discrimination for frequency-multiplexed optical BSM \cite{Sinclair} at the midpoint in this meet-in-the-middle scheme. This scheme can be applied to both one-photon entanglement \cite{Simon,Duan} and two-photon entanglement; in \cite{Sinclair}, there is no restriction on the degrees of freedom, but they should be consistent with frequency multiplexing. For example, the photon number is used for one-photon entanglement, and polarization and time bins are used for two-photon entanglement.

Let $p$ be the probability that a photon can reach the BSM site from each of the four entangled photon sources (EPSs) placed at Alice, Bob, and the repeater node depicted in Fig. \ref{fig:scheme}. $N$ is the number of frequency modes, and $p$ is assumed to be small enough that $(1-p)^N\sim1-Np$ holds.
First, in the case of a single mode, the probability that the photons can arrive from both sides, the measurement succeeds, and the entanglement can be shared by the elementary link is expressed as $p^2$. In the case of $N$ modes, the probability of successful BSM in at least one mode is $1-(1-p^2 )^N\sim Np^2$, which is a linear improvement. Next, let us consider two elementary links, $N$-mode multiplexing, and the BSM at the repeater node. Assuming that one mode succeeds on both the elementary links, the probability of successful BSM within the repeater node is expressed as follows:
\begin{equation}
P_{2,N}=\eta\times[1-(1-p^2 )^N ]^2\times1/N \sim Np^4  .
\end{equation}
Then the term $\eta$ incorporates the detection probability of the detector, efficiency of the QM, and probability of the successful BSM. The reason for the multiplier $1/N$ here is that success in the same mode is required in the two elementary links. However, if it is possible to perform frequency-mode matching, as proposed in Ref. \cite{Sinclair}, in which the repeater adjusts to the successful mode of the BSM in the elementary link, then
\begin{equation}
P_{2,N,match}=\eta\times[1-(1-p^2 )^N ]^2\sim N^2 p^4  ,
\end{equation}
and it is possible to obtain a nonlinear improvement in $N^2$ from the time of one mode.

On the other hand, owing to the advantage of not having to wait for two photons to arrive, schemes utilizing one-photon interference have been proposed \cite{Simon,Duan}. Furthermore, a significant experimental report \cite{Dario} on quantum communication using one-photon interference described the generation of entanglement between matter quantum bits (qubits) using EPSs and QMs with the AFC. One-photon interference can also benefit from $N$-mode multiplexing in the same way as two-photon interference. First, when focusing on a single elementary link, the probability of success in the one-mode case is $1-(1-p)^2$. When $N$-mode multiplexing is used, a total of $2N$ modes are sent to the interference system, so the probability of success is $1-(1-p)^{2N}\sim 2Np$, and a linear improvement is expected. Next, as before, we consider two elementary nodes: if there is no mode matching, then
\begin{equation}
P_{1,N}=\eta' \times [1-(1-p)^{2N} ]^2 \times 1/N  \sim 4Np^2  ,
\end{equation}
and if mode matching is possible, then
\begin{equation}
P_{1,N,match}=\eta' \times [1-(1-p)^{2N} ]^2 \sim 4N^2 p^2
\end{equation}
is obtained. $\eta'$ is the corresponding inefficiency in the case of single-photon interference. Therefore, both two-photon and one-photon interferences are improved by $N$-mode multiplexing, with a linear improvement without mode matching and a second-order improvement with mode matching. 

However, one-photon interference requires phase locking because the phases must be aligned in the two optical paths. Although phase locking is not easy, twin-field QKD, which requires phase locking or monitoring as well as one-photon interference-based entanglement distribution, has been demonstrated at distances of over 100 km \cite{Pittaluga, Chen}. In the demonstration, fiber phase control uses a high-intensity reference light that monitors the relative phase difference, and because the reference light is a noise source (Rayleigh scattering is the main factor), a method that separates either the signal photon and reference transmission \cite{Guo} in the time domain or the signal photon wavelength and reference light wavelength \cite{Pittaluga} is used. On the other hand, there are implementations that avoid relative phase control by devising the QKD protocol \cite{Chen}, but this is a unique technique for QKD and cannot be easily adapted to entanglement distribution. Based on scalability and technical cost, this paper mainly adopts two-photon interference because it is possible to obtain high entanglement generation rates using frequency multiplexing with two-photon interference \cite{Sinclair}. Finally, when on-demand QM is incorporated, $N$-mode multiplexing is expected to contribute to an increase in the BSM success rate and to shorten the required QM storage time. Therefore, multiplexing is useful in all cases.
\begin{figure*}
    \centering
    \includegraphics[width = 16cm]{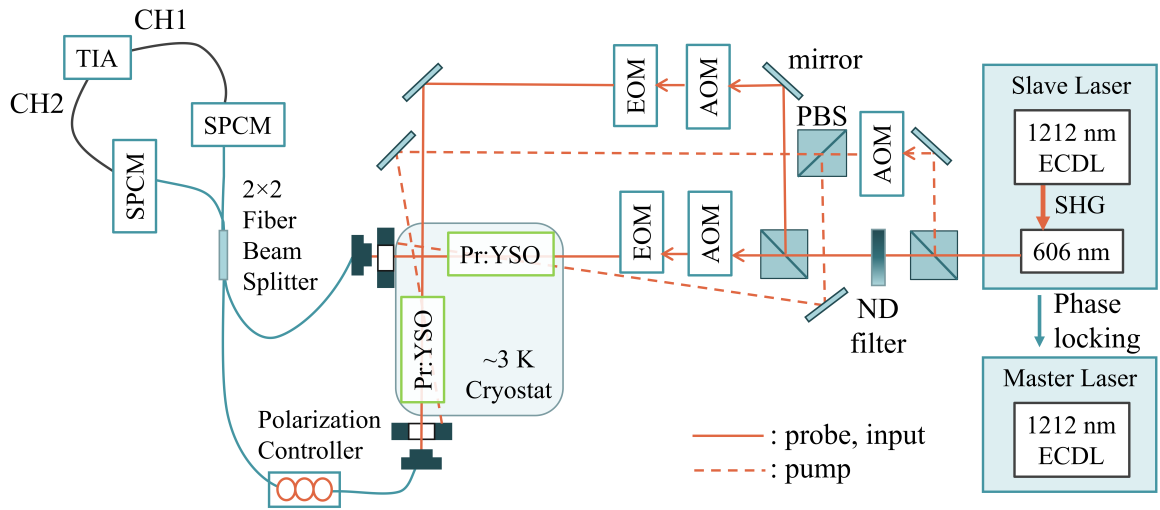}
    \caption{
     Experimental setup. Second harmonic generation (SHG) of 1212 nm external cavity diode laser (ECDL) was used. The laser used was phase locked to the master laser, which is frequency stabilized to an optical frequency comb, and the linewidth was of the order of 0.1 MHz. The laser beam was split into pump and probe beams for the two crystals by the polarizing beam splitter (PBS), and the probe beam was modulated by a neutral-density (ND) filter to a single-photon level before being injected into the three AFCs by an acousto-optic modulator (AOM) and electro-optic modulator (EOM). The AOMs are responsible for generating signal pulses and varying the time delay between the two paths. The photon echoes retrieved from each Pr:YSO with a concentration of 0.05\% were coupled to a single-mode fiber and interfered with by a 2 $\times$ 2 fiber beam splitter (FBS); following the FBS, the photons were observed by a single-photon detector, and the time difference between the two single-photon-counting modules (SPCMs) and the number of photons was recorded by the time-interval analyzer (TIA). Here, half-wave plates, quarter-wave plates, and lenses are omitted to avoid confusion.}
    \label{fig:setup}
\end{figure*}

%%%%%%%実験系！！！！%%%%%%%
\section{\label{level1-3}EXPERIMENTAL SETUP}

As illustrated in Fig. \ref{fig:setup}, the experimental setup consists of two Pr:YSOs cooled to approximately 3 K in a closed-cycle cryostat, a laser source (a pump light for the AFC preparation and a probe light for observation or storage for input in the AFC) and a HOM interferometer. A laser beam with a wavelength of 605.9773 nm was phase locked to an optical frequency comb locked to an absolute frequency \cite{Miyashitasan, Mannamisan, Kondosan}. Three AFCs were prepared and separated by 100 MHz \cite{Ortumulti}, each with a bandwidth of approximately 10 MHz and comb spacings $\Delta$ of 1.53, 0.92, and 0.657 MHz (with storage times of 0.652, 1.09, and 1.52 {\textmu}s), respectively. The preparation time for all three AFCs was 420 ms. The probe light was modulated by an acousto-optic modulator (AOM) to Gaussian pulses with $t_{FWHM} = 100$ ns, and then sidebands were generated by an electro-optic modulator (EOM) driven by a 100-MHz sine wave for frequency multiplexing the three modes: negative first, zeroth, and first orders. Therefore, input beams were prepared with three frequency components that matched the three AFCs generated every 100 MHz. Each time, frequency-multiplexed AFCs were generated, and 45 000 frequency-multiplexed Gaussian pulses (450 ms) were input. In this case, a precise modulation-and-locking system was constructed using the dynamic phase-lock technique \cite{Numata}. The light entered each crystal through a convex lens that was focused onto the Pr:YSO. To observe echoes at the classical light level, the light power was monitored by a photodetector (PD). The beam diameters were 500 and 100 {\textmu}m, respectively, for the pump and probe at the centers of the Pr:YSOs.

\begin{figure*}
    \centering
    \includegraphics[width = 16cm]{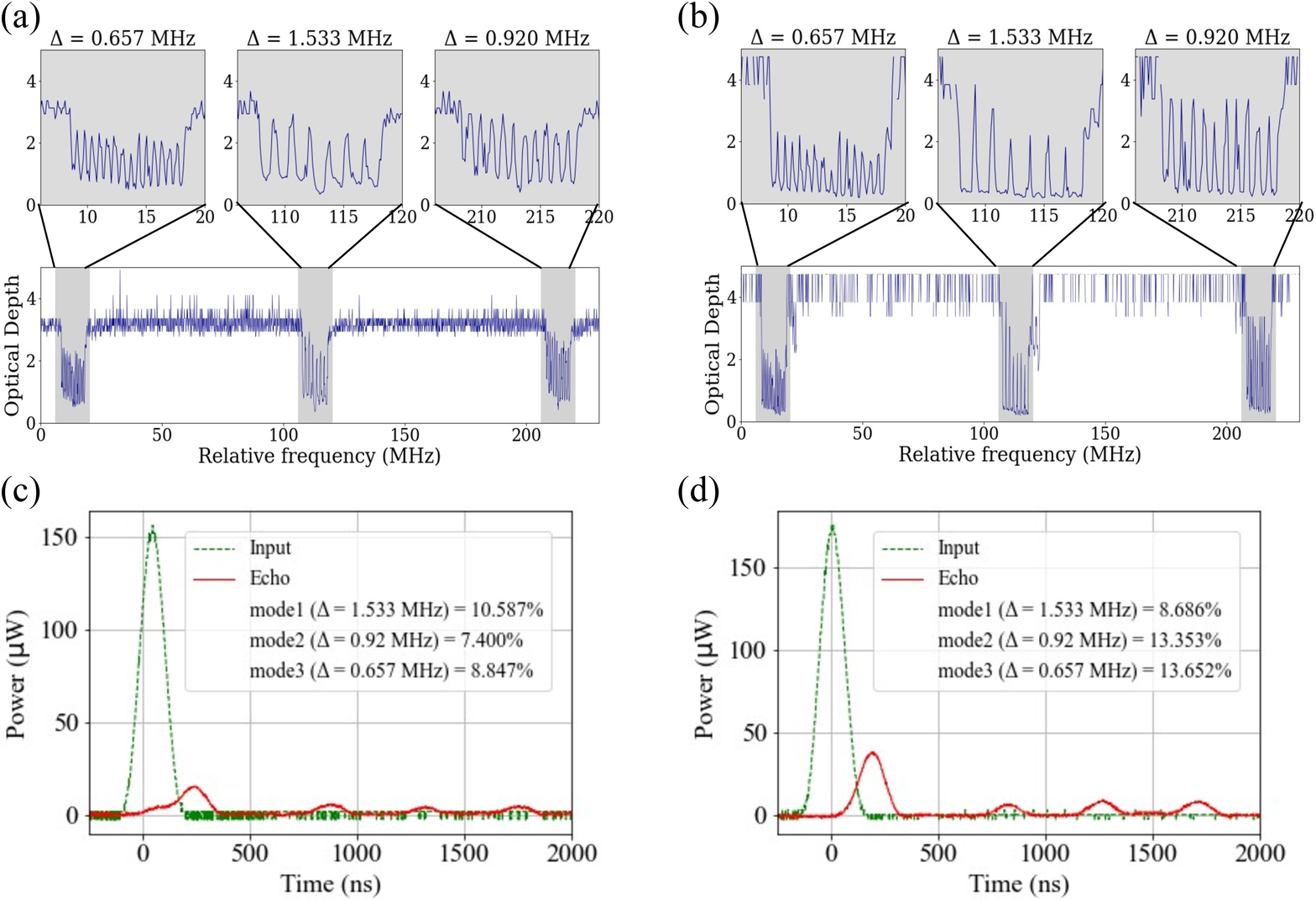}
    \caption{
     Observed AFCs and photon echoes in the classical level. (a) and (c) depict the results for crystal 1, and (b) and (d) depict those for crystal 2; the peak at approximately 0 ns is the transmitted light that was not stored in the AFC.}
    \label{fig:AFC}
\end{figure*}

Photons retrieved from the three AFCs with various comb spacings in each Pr:YSO were separated in time and coupled by a single-mode fiber before being sent to a HOM interferometer consisting of a fiber beam splitter (FBS). The polarization controllers were set on one side to ensure that the two input photons had the same polarization. Because the fiber length was not locked, we can say that the phases of the two lights can be randomized. Moreover, because two Pr:YSOs have different AFC retrieval efficiencies and fiber-coupling efficiencies, care must be taken to align the power when the photons enter the FBS to obtain maximum HOM interference. The two FBS outputs were each sent to a single-photon detector (SPD). The time difference of photon detections by SPDs was measured with a time-interval analyzer (TIA) to obtain the probability of coincidence. In this experiment, the HOM dip was measured by varying the delay time in the input channel; when the time difference between the inputs provided to the two crystals is zero, the probability of coincidence should be the lowest because the indistinguishability of the two photons is the highest. The time difference between sending the input light to channel 1 (corresponding to crystal 1) and channel 2 (crystal 2) was varied by 20 ns in the range from -500 to +500 ns.

%%%%%%%実験結果！！！！%%%%%%%
\section{\label{level1-4}RESULTS AND DISCUSSION}
\subsection{\label{level2-1}AFCs and photon echoes}
First, we report the observations of the generated AFCs and echoes in the classical level. Figures \ref{fig:AFC}(a) and \ref{fig:AFC}(b) depict the AFCs and Figs. \ref{fig:AFC}(c) and \ref{fig:AFC}(d) depict the echo created in the respective crystals. Notably, Figs. \ref{fig:AFC}(a) and \ref{fig:AFC}(c) were created with crystal 1, and Figs. \ref{fig:AFC}(b) and \ref{fig:AFC}(d) were created with crystal 2. It can be seen that the three AFCs with different comb spacings were generated every 100 MHz through direct laser modulation. The values of the comb spacing $\Delta$ are 0.657, 1.53, and 0.92 MHz. The comb spacing is determined from the Pr:YSO transition frequencies to create an AFC with a bandwidth of 9.2 MHz. The efficiency of the echoes was approximately 7.4\%--13.7\%. The intensity ratio of the input signal (probe laser) split into three bands (negative first, zeroth, and first orders) was approximately 1:1:1 by adjusting the applied voltage of the EOM. The photon echo efficiency is normalized by the input light intensity divided by 3. Therefore, the input light is split by the BS in front of the crystal and observed by the monitor PD to estimate the actual power from the split ratio. The photon echo efficiency can be further improved; however, a high efficiency generates a second echo owing to the re-absorption of echoes, and this causes an extra peak when we consider the HOM interference. If the second echo could be temporally excluded using an AOM shutter or other means, interference between more efficient echoes would be possible.
\begin{figure*}
    \centering
    \includegraphics[width = 16cm]{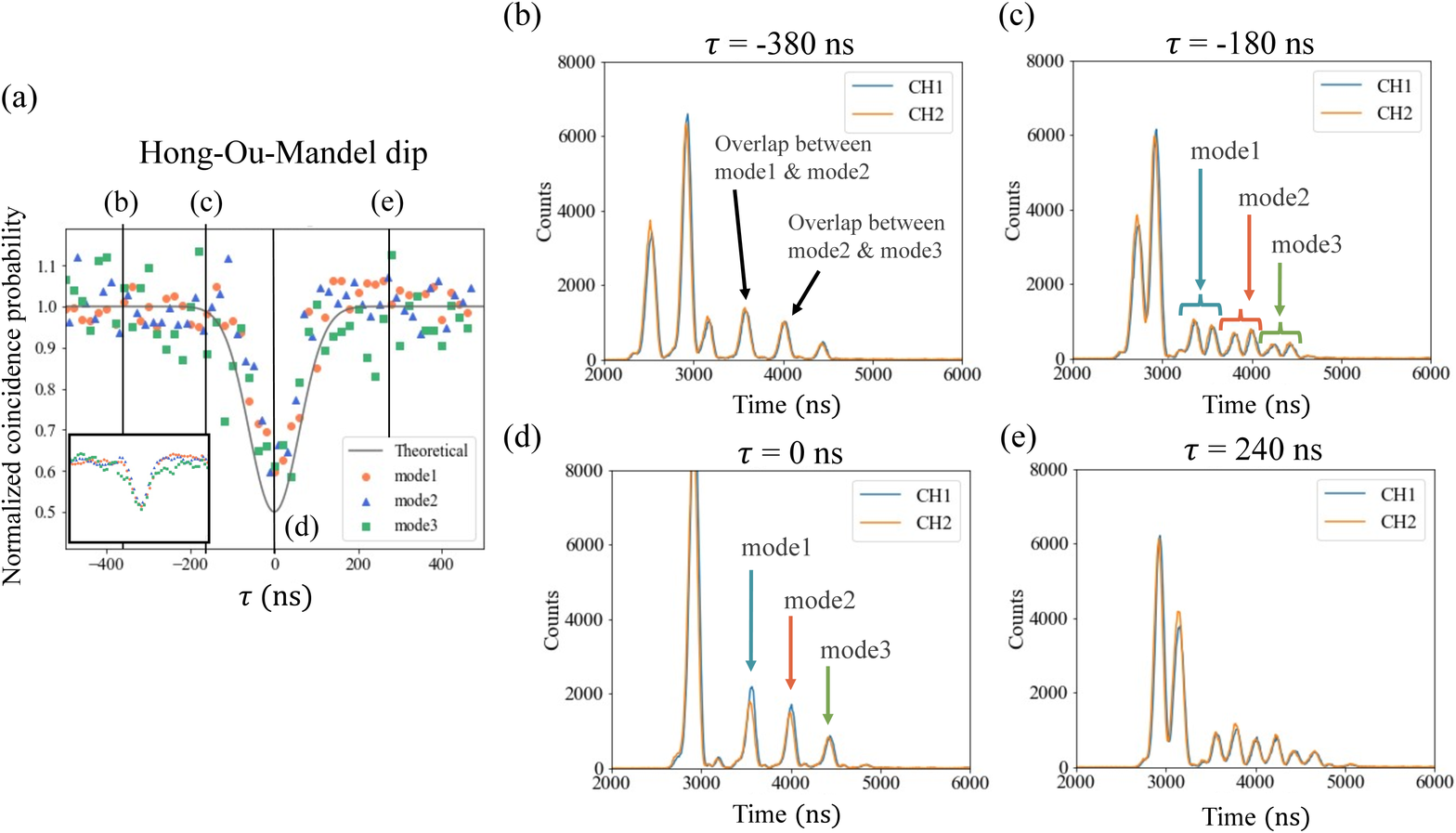}
    \caption{
     (a) Observed Hong-Ou-Mandel dip  (for clarity, error bars are shown); the circles, triangles and squares represent the results for the first, second, and third frequency modes, respectively. The curve represents the theoretical dip. The horizontal and vertical axes denote the time difference between the two inputs and the normalized coincidence detection rate, respectively. The visibility is over 40\% in every case. (The inset is the adjacent average over five points.) (b)--(e) Histograms at specific time differences $\tau$. The count rate on the TIA at a given time difference on the horizontal axis is represented on the vertical axis. The bin size is 8.192 ns. The peak around 2000--3000 ns is transmitted light that was not stored in the AFC and has no physical significance. (b) $\tau$ = -380 ns, where mode 2 of crystal 1 overlaps mode 1 of crystal 2 and mode 3 of crystal 1 overlaps mode 2 of crystal 2. (c) $\tau$ = -180 ns and (e) $\tau$ = 240 ns. In this case, there is no overlap between the modes. (d) $\tau$ = 0 ns, at which all the modes overlap, and the deepest HOM dip of is expected to appear. At each time instant, the highest peak is the input light transmission.}
    \label{fig:dip}
\end{figure*}

\subsection{\label{level2-2}Hong-Ou-Mandel dip}
Figure. \ref{fig:dip}(a) presents the HOM interference results. The circles, triangles and squares represent the interference results between the echoes at $\Delta$ = 1.53, $\Delta$ = 0.92, and $\Delta$ = 0.652 MHz, respectively. According to \cite{Agata}, the theoretical dip for Gaussian input light is expressed as follows:
\begin{equation}
P_{coin}=1-\frac{1}{2} \exp⁡(-\frac{\sigma^2 \tau^2}{2})
\end{equation}
where $\sigma$ is $\sqrt{2}$ times the standard deviation of the spectral widths of input signals and $\tau$ is the time difference between the two inputs. The black curve in Fig. \ref{fig:dip}(a) represents the theoretical dip obtained from the simulation based on Eq. (6). According to Fig. \ref{fig:dip}(a), the data, especially for mode 3, are noisy. Given that the visibility depends on the average photon number of laser (Poisson) signal photons, the average photon number was kept low to obtain high HOM interference, resulting in a relatively high noise due to a large contribution of the dark count of the SPDs. 

In addition, to clearly see the HOM dip, the adjacent mean of five points is also depicted. Because we used weak coherent states, the visibility of the HOM dip is limited to 50\% owing to the Poisson distribution of the laser. The visibility in this experiment was 40\%--42\% for every case. There are two causes of visibility degradation: (i) lower indistinguishability owing to fluctuations in the power ratio and (ii) experimental environmental factors such as the detection efficiency of SPDs and average photon number. To start, let us discuss the first case. Figures. \ref{fig:dip}(b)--\ref{fig:dip}(d) present  histograms representing two channels for each value of delay time $\tau$. The ratios of echo intensities from the same frequency mode in the two crystals differ due to variation in laser power caused by fluctuations in polarization. Because HOM interference is a phenomenon that occurs between two indistinguishable photons, the difference between the two input wave packets affects the visibility. Therefore, a slight fluctuation of the echo intensity from the same frequency mode read from the histogram [especially Figs. \ref{fig:dip}(c) and \ref{fig:dip}(e)] can be a cause of visibility degradation. 

Further, we confirm the relation between visibility and the actual experimental conditions such as the detection efficiency of SPDs and average photon number in the second cause of visibility degradation. According to Ref. \cite{Ele}, visibility is expressed as follows:
\begin{equation}
V=1-\frac{P^{(coin)}}{P^{(c)} P^{(d)}}
\end{equation}
when
\begin{multline}
P^{(coin)}=1-CI_0(2\eta_c \sqrt{\mu_a \mu_b} tr \cos \phi)\nonumber\\
-DI_0 (\eta_d \sqrt{\mu_a \mu_b} tr \cos \phi )\\\nonumber
+CD I_0 (2(\eta_c-\eta_d ) \sqrt{\mu_a \mu_b }tr \cos \phi )),\nonumber
\end{multline}
\begin{equation}
\begin{cases}
C=\exp⁡[-\eta_c (\mu_a t^2+\mu_b r^2 )](1-d_c),\nonumber \\
D=\exp⁡[-\eta_d (\mu_a t^2+\mu_b r^2 )](1-d_d), \nonumber \\
\end{cases}
\end{equation}
\begin{equation}
\begin{cases}
P^{(c)}=1-CI_0 (2\eta_c \sqrt{\mu_a \mu_b} tr \cos \phi),\nonumber \\
P^{(d)}=1-DI_0 (2\eta_d \sqrt{\mu_a \mu_b} tr \cos \phi).\nonumber
\end{cases}
\end{equation}

\begin{figure*}
    \centering
    \includegraphics[width = 16cm]{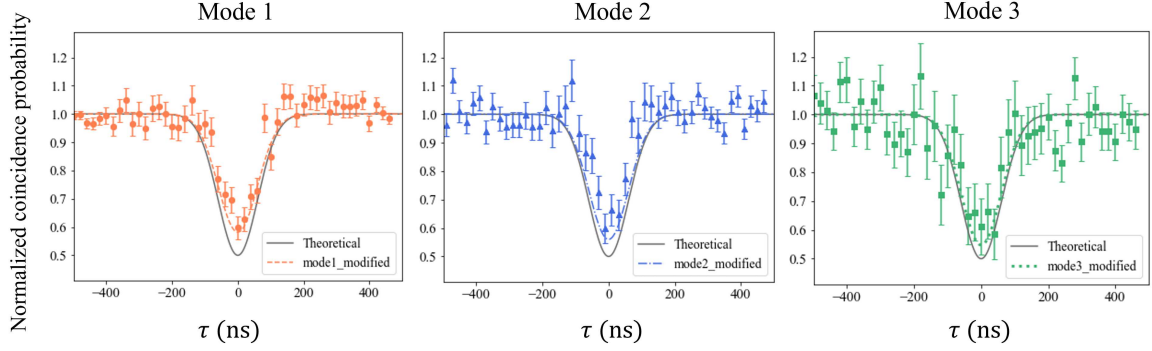}
    \caption{
     Comparison of experimental and simulation results, considering the two visibility degradation factors. All error bars are calculated as the square root of the total counts. When the modified simulations are compared with the solid curve (the theoretical value without modification incorporating the factors), it can be seen that the simulation and experimental values are closer.}
    \label{fig:ModiDip}
\end{figure*}
where $I_0$ is the zeroth-order deformed Bessel function of the first kind, $\mu_a$ and $\mu_b$ are the average photon numbers for each input, $\eta_c$ and $\eta_d$ are the detection efficiencies of the SPDs, $d_c$ and $d_d$ are the dark-count detection probabilities of the SPDs, $t$ and $r$ are the square roots of the transmittance and reflectance of the BS used, and $\phi$ is the polarization shift between the two inputs. $\mu,\eta,d,$ and $\phi$ are measured values, and in this experiment, $\mu_a=\mu_b=$ 0.064 (mode 1), 0.053 (mode 2), and 0.031 (mode 3), $\eta_c=\eta_d=0.47$, $d_c=d_d=1.5\times10^{-4}, t=r=1/\sqrt{2},$ and $\phi=3\tcdegree$. Parameters except $\mu_a$ and $\mu_b$ were determined by the equipment used, and $\mu_a$ and $\mu_b$ were chosen to have the largest $V$ with reference to \cite{Ele}. With these parameters, the visibility limit was found to be 48.6\% for mode 1, 47.2\% for mode 2, and 47.1\% for mode 3. 

\begin{table}[h]
\caption{\label{tab1}%
Change in the visibility by a modification incorporating the degradation factors. The values were obtained from fitting Eq. (6).}
\begin{ruledtabular}
\begin{tabular}{lccc}
\multicolumn{1}{c}{\textrm{}}&
1.533 MHz&
0.92 MHz&
0.652 MHz\\
\colrule
Experiment & 42\% & 42\% & 40\%\\
Modified by only (i) & 43\% & 46.5\% & 48\%\\
Modified by (i) and (ii) & 41\% & 44\% & 45\%\\
\end{tabular}
\end{ruledtabular}
\label{tb:dip}
\end{table}

Therefore, considering the two factors that degrade the visibility, the modified dip incorporating the factors is shown in Fig. \ref{fig:ModiDip}. Table \ref{tb:dip} presents a summary of the visibility. It can be seen that the experimental and simulated visibilities are closer when both the causes of degradation factors are considered, compared with the case when only the first cause is considered. However, there is still a difference between the experimental and theoretical results at 0.652 MHz. This can be attributed to the low efficiency because of the narrow $\Delta$ \cite{Afzelius} in the third echo and the small number of interference trials, as seen in Figs. \ref{fig:dip}(c) and \ref{fig:dip}(e). A decrease in visibility leads to a decrease in fidelity in the BSM. To prevent this, the use of more stable optical systems and SPDs with higher detection efficiency and lower dark counts are required.

Further, we show that there is no mixing with another mode. In Fig. \ref{fig:dip} (b), mode 2 of crystal 1 overlaps with mode 1 of crystal 2, and mode 3 of crystal 1 overlaps with mode 2 of crystal 2. Therefore, if there is mixing between the different modes here, interference would occur, and the coincidence rate would be reduced. However, because no extra dip is seen at the edge of the dip in Fig. \ref{fig:dip}(a), it can be said that there is no interference among the different modes and that the modes are well separated. Since there is no mode mixing, the developed technique leads to frequency-multiplexed BSM.

\subsection{\label{level2-3}Application to quantum repeater}
In this section, we will discuss how the results of this experiment can be applied to our scheme. In future studies, this experiment can be used to implement optical BSMs with frequency multiplexing and to generate a heralding signal for the quantum repeater. The frequency (wavelength) channel information from the optical BSM is converted into time information by the AFC, a heralding signal generated by SPDs consisting of a BSM returned to the repeater through the classical communication channel. The repeater obtains the information about the frequency channel stored in the QM by reading the heralding signal which informs the frequency (wavelength) channel from the time information and regenerates the signal photons of that channel. Therefore, to implement the procedure suggested in Ref. \cite{Sinclair}, a frequency shift is applied before entanglement swapping is carried out. The problem is that the frequency (wavelength) information cannot be obtained unless the time of photon generation at the entangled photon source is known. However, this problem can be overcome by combining the time-to-space and frequency-to-time mode mappings \cite{Yoshidasan}.

Here, we consider the upper bound of the heralding rate calculated from the echo’s retrieved time in this experiment. The frequency-to-time mode mapping carried out by the AFC sets an upper bound on the rate at which the photons can be transmitted because a single temporal (frequency multiplexed) mode is now separated into multiple temporal modes owing to frequency-to-time mode mapping. The upper limit of the received photon rate $R_{limit}$ at the optical BSM can be calculated using the number of modes $N$, the maximum AFC storage time $t_{AFC}$, AFC efficiency $\eta_{AFC}$, fiber transmission $L_{fiber}$, and wavelength conversion efficiency $R_{WC}$. Here, we assume that photons are transmitted for 100 km along a telecommunication wavelength band with low transmission loss. Considering the need to convert the telecommunication wavelength band to a wavelength suitable for QM, two-photon interference, and one elementary link, with $N$ = 3, $t_{AFC}$ = 1.52 {\textmu}s, $\eta_{AFC}$ = 0.1, $L_{fiber}$  = 0.1 (at 50 km for two elementary links), and $R_{WC}$  = 0.5 \cite{Niizekisan}, we obtain the following equation:

\begin{equation}
R_{limit}=\frac{N}{t_{AFC}} \eta_{AFC}^2 L_{fiber}^2 R_{WC}^2=49 counts/s.
\end{equation}

Although an efficiency of 10\% was obtained for the AFC and only three-mode multiplexing was used in this study, the achievable value of the efficiency is 54\% \cite{Afzelius}. Increasing the number of modes will increase $t_{AFC}$ with only a slight increase of $R_{limit}$. However, the interval of input photons to BSM can be limited by using a time-to-spatial mode mapper \cite{Yoshidasan}. So it is possible to further improve $R_{limit}$ by increasing $\eta_{AFC}$ and $N$ while maintaining a short $t_{AFC}$.

For simplicity, we consider the case without the time-to-spatial mode mapper. We can obtain a feasible and optimum value of $3.2 \times 10^2$ counts/s when $N$=30 \cite{Yoshidasan}, $t_{AFC}$=13.3 {\textmu}s, $\eta_{AFC}$=0.2, and $R_{WC}$=0.6 \cite{Niizekisan}. In Ref. \cite{Xiao}, a heralding rate of 100 counts/s was obtained with a 10-m intralab fiber transmission with a rare-earth-doped crystal. Given the distances involved, the scheme considered in this study may be useful.

Next, we consider one-photon interference. In this case, for the same calculation as in (8), $R_{limit}$ is given by

\begin{equation}
R_{limit}=\frac{N}{t_{AFC}} \eta_{AFC} L_{fiber} R_{WC}=9.9 \times 10^3 counts/s
\end{equation}
and feasible optimum $R_{limit}=2.7 \times 10^4$ counts/s. Recent studies on memory entanglement showed a heralding rate of 1.43 kilocounts/s (with 50-m fiber) for rare-earth-doped crystals \cite{Dario}. In a recent study of twin-field QKD \cite{Pittaluga}, the key generation rate was $\sim$$4 \times 10^4$ bits per second (bps) for a 150-km transmission and $\sim$40 bps for a 500-km transmission. This is close to the assumed value for one-photon interference using our system, but it shows that a better key generation rate is possible with the assumed optimal value. Moreover, since this research involves frequency multiplexing of optical BSM leading to entanglement generation, it has advantages beyond key generation by QKD. This study shows that much larger heralding rates can be obtained even at longer distances.

Finally, a comparison with previous studies is made: in \cite{Jeongwan}, HOM interference was observed between retrieved photons from two Ti:Tm:$\mathrm{LiNbO_3}$ waveguides. The signal photons have the same single frequency mode with a wide bandwidth of a 600-MHz AFC. The visibility is as high as 47.9\% $\pm$ 3.1\%. We observed HOM interference in the retrieved photons from multiple frequency modes. In \cite{Oriol}, input pulses consisting of multiple frequency modes were separated by a virtually imaged phased array (VIPA). It allows the selection of frequency bands and does not require a cooler, but it is difficult to achieve a resolution as high as 100 MHz with the use of a VIPA. The scheme for frequency-multiplexed entanglement distribution may vary depending on the type of QM and photon source used in the quantum repeater. A comparison of communication schemes as a whole will be carried out in future studies.

%%%%%%%結論！！！！%%%%%%%
\section{\label{sec:level1}CONCLUSION}
In conclusion, we have successfully established HOM interference with frequency multiplexing. Three AFCs with different comb spacings were created every 100 MHz, and the Hong-Ou-Mandel interference was observed between the photon echoes retrieved at different times from each of the AFCs. The visibility is above 40\% in every case. Further, as there is no mixing between the different modes, it can be said that indistinguishability is maintained for the photons in every mode. Because the inhomogeneous broadening of Pr:YSO is of the order of 10 GHz and the AFC bandwidth is of the order of 10 MHz, it is possible to further increase the number of modes. This experiment indicates the possibility of realizing a quantum repeater with a high entanglement generation rate by frequency-multiplexed BSM in the future.

\begin{acknowledgments}
This research was supported by the SECOM Foundation, JSPS KAKENHI (Grant No. JP20H02652), NEDO (Grant No. JPNP14012), and JST Moonshot R\&D Grant No. JPMJMS226C. We also acknowledge the members of the Quantum Internet Task Force, which is a research consortium to realize the quantum internet, for comprehensive and interdisciplinary discussions of the quantum internet.
\end{acknowledgments}

\bibliography{main}% Produces the bibliography via BibTeX.

\end{document}